
\magnification \magstep1
\raggedbottom
\openup 4\jot
\voffset6truemm
\headline={\ifnum\pageno=1\hfill\else
\hfill {\it Cosmological restrictions on
conformally invariant $SU(5)$ GUT models}\hfill
\fi}
\centerline {\bf COSMOLOGICAL RESTRICTIONS ON CONFORMALLY INVARIANT}
\centerline {\bf $SU(5)$ GUT MODELS}
\vskip 1cm
\centerline {\bf Giampiero Esposito$^{1,2}$,
Gennaro Miele$^{3,4}$ and Luigi Rosa$^{3,4}$}
\vskip 1cm
\centerline {$^{1}${\it International Centre for Theoretical Physics}}
\centerline {\it Strada Costiera 11, 34014 Trieste, Italy;}
\centerline {$^{2}${\it Scuola Internazionale Superiore di Studi Avanzati}}
\centerline {\it Via Beirut 2-4, 34013 Trieste, Italy;}
\centerline {$^{3}${\it Dipartimento di Scienze Fisiche}}
\centerline {\it Mostra d'Oltremare Padiglione 19, 80125 Napoli, Italy;}
\centerline {$^{4}${\it Istituto Nazionale di Fisica Nucleare}}
\centerline {\it Mostra d'Oltremare Padiglione 20, 80125 Napoli, Italy.}
\vskip 5cm
\centerline{SISSA Ref. 201/92/A (October 1992)}
\vskip 100cm
\noindent
{\bf Abstract.} Dirac's theory of constrained Hamiltonian systems is
applied to the minimal conformally-invariant $SU(5)$ grand-unified model
studied at $1$-loop level in a de Sitter universe.
For this model, which represents a simple and interesting example of GUT
theory and at the same time is a step towards theories with larger gauge
group like $SO(10)$, second-class constraints in
the Euclidean-time regime exist.
In particular, they enable one to prove that,
to be consistent with the experimentally
established electroweak standard model
and with inflationary cosmology, the residual gauge-symmetry group
of the early universe, during the whole de Sitter era, is bound to be
$SU(3) \times SU(2) \times U(1)$.
Moreover, the numerical solution of the field equations subject
to second-class constraints is obtained. This confirms the
existence of a sufficiently long de Sitter phase of the early universe,
in agreement with the initial assumptions.
\vskip 8cm
\leftline {PACS numbers: 02.60.Cb, 04.20.Fy, 11.15.Ex, 98.80.Cq,
98.80.Dr}
\vskip 100cm
\leftline {\bf 1. Introduction}
\vskip 1cm
\noindent
In recent work by the first two authors (Buccella {\it et al}
1992) the spontaneous-symmetry-breaking pattern of $SU(5)$ gauge
theory was studied in a de Sitter universe. The main result was
the proof that the technique described in Buccella {\it et al}
1980 to study spontaneous symmetry breaking of $SU(n)$ for
renormalizable polynomial potentials in flat spacetime can be
generalized, for $SU(5)$, to the curved background relevant to
the inflationary cosmology (i.e. de Sitter). One thus obtained a
better understanding of the result, previously found with a
different numerical analysis (Allen 1985), predicting the slide
of the inflationary universe into either the
$SU(3) \times SU(2) \times U(1)$ or $SU(4) \times U(1)$
extremum.

The main tool used was the Wick-rotated path integral for
Yang-Mills-Higgs theory at $1$-loop level about curved
backgrounds (leading to a de Sitter model with $S^{4}$
topology), and the corresponding $1$-loop effective
potential $V(r,{\bf {\hat \Phi}})$ first derived in Allen 1985.
Assuming that the Higgs field ${\bf {\hat \Phi}}={\rm diag}
\Bigr(\varphi_{1},\varphi_{2},
\varphi_{3},\varphi_{4},\varphi_{5}\Bigr)$
belongs to the adjoint representation of $SU(5)$, and using the
bare Lagrangian and the tree potential appearing in equations
(2.1)-(2.2) of Buccella {\it et al} 1992, the background-field
method and the choice of 't Hooft's gauge-averaging term lead
to (Allen 1985, Buccella {\it et al} 1992)
$$ \eqalignno{
V(r,{\hat {\bf \Phi}}) &= {15\over 64 \pi^{2}}
\left \{Q+{1\over 3} \Bigr(1-\log(r^{2}M_{X}^{2})\Bigr)
\right \} R \; g^{2} {\parallel {\hat {\bf \Phi}} \parallel}\cr
&+ \left \{ {9\over 128 \pi^{2}}
\Bigr(1-\log(r^{2}M_{X}^{2})\Bigr)
-{21\over 320 \pi^{2}} \widetilde{\Lambda} \right \}
g^{4}{\parallel {\hat {\bf \Phi}} \parallel}^{2}\cr
&+ {15\over 128 \pi^{2}}
\left \{ {12\over 5} \widetilde{\Lambda} +
\Bigr(1-\log(r^{2}M_{X}^{2})\Bigr) \right \}
g^{4}\sum_{i=1}^{5}{\varphi_{i}}^{4}\cr
&- {3\over 16\pi^{2}r^{4}}\sum_{i,j=1}^{5}
{\cal A}\left[{r^{2}g^{2}\over 2}
{(\varphi_{i}-\varphi_{j})}^{2} \right] + V_{0}
&(1.1)\cr}
$$
where $Q$ has been defined in equation (2.6) of Buccella {\it et al}
1992, and we here denote by ${\widetilde \Lambda}$ the parameter
defined in equation (2.7) of Buccella {\it et al} 1992, to avoid
confusion with the cosmological constant $\Lambda$. Note also that
${\cal A}$ denotes the special function defined in equation (A.1)
of the appendix,
$r = \sqrt{3 / \Lambda}$ is the four-sphere radius, and the
constant $V_{0}$ is equal to the desired
$V({\bf \Phi}=0)$ value.

A naturally-occurring question is whether further restrictions on
the $SU(5)$ broken-symmetry phases can be derived within the
framework of inflationary cosmology. This paper is devoted to the
study of such a problem, and is thus organized as follows.
Section 2 performs the Hamiltonian analysis of the Riemannian
(i.e. Euclidean-time) version of a de Sitter background coupled
to the $SU(5)$ model, following Dirac's theory of constrained
systems. Section 3 studies the $1$-loop effective potential
in the $SU(3) \times SU(2) \times U(1)$ and
$SU(4) \times U(1)$ broken-symmetry phases. Section 4 shows
that to be compatible with the
prescriptions of the electroweak standard
model and with the inflationary scheme,
the early universe  can only reach the
$SU(3) \times SU(2) \times U(1)$ broken
phase and that during all the de Sitter
era, possible tunneling processes
towards the $SU(4) \times U(1)$ invariant phase are
energetically forbidden. Section 5 describes the numerical
integration of the corresponding field equations for the
four-sphere radius and the components of the Higgs field.
Concluding remarks are presented in section 6.
\vskip 10cm
\leftline {\bf 2. Hamiltonian analysis}
\vskip 1cm
\noindent
The effective Lagrangian for the
Riemannian version of an exact de Sitter
background with $S^{4}$ topology coupled to the $SU(5)$ model
is given by (up to a multiplicative constant)
$$
L = \left[-{3 r^2 \over 4 \pi G } + r^4 \left( \sum_{i=1}^{5}
{\dot{\varphi}^{2}_{i} \over 2}
+ V(r,{\hat {\bf \Phi}}) \right) \right]
\eqno (2.1)
$$
where the derivatives ${\dot \varphi}_{i}$ are taken with respect
to the Euclidean time $\tau$ and the
spatial gradient of the Higgs field has been assumed to
be negligible. Note that, after having integrated
over the gauge-bosons degrees of freedom, the effective potential
only involves the Higgs field.
According to the simplified argument
usually presented in the literature, the field equations obtained
by varying the action with respect to $r$ and ${\hat {\bf \Phi}}$
are then (Allen 1983)
$$
r^2 = {3 \over 8 \pi G \left[ \sum_{i=1}^{5}
{ \dot{\varphi}^{2}_{i} / 2} + V(r,{\hat {\bf \Phi}}) \right]}
\eqno (2.2a)
$$
$$
\ddot{\varphi}_{i}  = {\delta V(r,{\hat {\bf \Phi}}) \over
{\delta \varphi_{i}}}~~~~~~~\forall i=1,..,5~~~.
\eqno (2.2b)
$$
However, this approach does not take into account the full Hamiltonian
treatment of the problem. In other
words, from equation (2.1) one derives
the primary constraint $p_{r} \approx 0$, where $p_{r}$ is the
momentum conjugate to the four-sphere
radius, and $\approx$ is the symbol
of weak equality (i.e. an equality which only holds on the
constraint surface).
This primary constraint should be preserved using the
technique described for example in Dirac 1964 and Esposito 1992. The
corresponding Hamiltonian analysis is as follows.

The canonical Hamiltonian $H_{c}$, defined as the Legendre transform of
the Lagrangian $L$ in equation (2.1), takes the form
$$
H_{c}= r^{-4} \sum_{i=1}^{5} {p_{\varphi_{i}}^{2} \over 2} + b r^2 -
r^{4} V(r,{\hat {\bf \Phi}})
\eqno (2.3)
$$
where we have defined $b \equiv 3/(4 \pi G)$. Thus, the effective
Hamiltonian ${\widetilde H}$ defined on the whole phase space becomes
$$
\widetilde{H} \equiv  H_{c} +
\lambda(r,{\hat {\bf \Phi}},p_{r},p_{{\hat {\bf \Phi}}})
{}~p_{r}~~~.
\eqno (2.4)
$$
The preservation of the primary constraint $V_{1} \equiv p_{r}$ yields
$$
\dot{p}_{r} \equiv \Bigr \{ p_{r}, \widetilde{H} \Bigr \}
\approx 2 r^{-5} \sum_{i=1}^{5} p_{\varphi_{i}}^{2} - 2 b r +
4 r^3 V(r,{\hat {\bf \Phi}}) + r^4 {\delta V \over \delta r}
\eqno (2.5)
$$
where $\Bigr \{ \; \; , \; \; \Bigr \}$ denote
the Poisson brackets.
One therefore finds in our model the secondary constraint
$$
V_{2} \equiv 2 r^{-5} \sum_{i=1}^{5} p_{\varphi_{i}}^{2} - 2 b r +
4 r^3 V(r,{\hat {\bf \Phi}}) + r^4 {\delta V \over \delta r}~~~.
\eqno (2.6)
$$
Further constraints are not found, since the preservation of $V_{2}$
leads to the condition
$$
0 = \dot{V}_{2} \equiv \Bigr \{ V_{2} , \widetilde{H} \Bigr \} \approx
\Bigr \{ V_{2}, H_{c} \Bigr \} + \lambda \Bigr \{ V_{2}, V_{1} \Bigr \}
\eqno (2.7)
$$
which can be solved for $\lambda$ as
$$
\lambda = - {\Bigr \{ V_{2}, H_{c} \Bigr \}  \over
\Bigr \{ V_{2}, V_{1} \Bigr \}  }~~~.
\eqno (2.8)
$$
This solution can be obtained since the constraints $V_{1}$ and
$V_{2}$ are second-class, and the Poisson brackets appearing in
the formula for $\lambda$ are found to be
$$ \eqalignno{
\Bigr \{ V_{2}, H_{c} \Bigr \} & =
-{4 \over r} \sum_{i=1}^{5} \Bigr \{ p_{\varphi_{i}}^{2},
V(r,{\hat {\bf \Phi}}) \Bigr \} - {1 \over 2}
\sum_{i=1}^{5} \biggr \{ p_{\varphi_{i}}^{2},
{\delta V \over \delta r} \biggr \}\cr
&={8 \over r} \sum_{i=1}^{5}
p_{\varphi_{i}} {\delta V \over \delta \varphi_{i}}
+ \sum_{i=1}^{5} p_{\varphi_{i}} {\delta^{2} V \over \delta \varphi_{i}
\delta r}
&(2.9)\cr}
$$
$$
\Bigr \{ V_{2},V_{1} \Bigr \} = - 10 r^{-6} \sum_{i=1}^{5}
p_{\varphi_{i}}^{2} - 2 b + 12 r^2 V(r,{\hat {\bf \Phi}}) +
r^4 {\delta ^{2} V \over \delta r^{2}}
+ 8 r^3 {\delta V \over \delta r}~~~.
\eqno (2.10)
$$
Moreover, since $V_{1}$ and $V_{2}$ are second-class, they can be set
strongly to zero using Dirac brackets (Dirac 1964, Esposito 1992),
hereafter denoted by $\left \{ \; \; , \; \; \right \}^{*}$. The
corresponding field equations are
$$
\dot{r} \approx \Bigr \{ r, \widetilde{H} \Bigr\}^{*} \approx \lambda
\eqno (2.11)
$$
$$
\dot{\varphi}_{i} \approx \Bigr \{ \varphi_{i}, \widetilde{H} \Bigr\}^{*}
\approx r^{-4} p_{\varphi_{i}} \left( 1 - {2 \over r}{V_{2} \over \{ V_{1},
V_{2} \} } \right)
\eqno (2.12)
$$
$$
\dot{p}_{r} \approx \Bigr \{ p_{r}, \widetilde{H} \Bigr\}^{*} \approx
\Bigr \{ p_{r}, H_{c} \Bigr \} - \Bigr \{ p_{r}, V_{l} \Bigr \}
C_{lm}^{-1} \Bigr \{ V_{m}, \widetilde{H} \Bigr \} \approx 0
\eqno (2.13)
$$
$$ \eqalignno{
\dot{p}_{\varphi_{i}} & \approx \Bigr \{ p_{\varphi_{i}}, \widetilde{H}
\Bigr\}^{*} \approx \Bigr \{ p_{\varphi_{i}}, H_{c} \Bigr \} -
\Bigr \{ p_{\varphi_{i}}, V_{l} \Bigr \}
C_{lm}^{-1} \Bigr \{ V_{m}, \widetilde{H} \Bigr \}\cr
& \approx r^4 {\delta V \over \delta \varphi_{i}} + \left(
4 r^3 {\delta V \over \delta \varphi_{i}}
+ r^4 {\delta^2 V \over
\delta r \delta \varphi_{i}} \right)
{V_{2} \over \{ V_{1},V_{2} \} }
&(2.14)\cr}
$$
where $C_{lm}$ is the matrix of Poisson brackets of second-class
constraints. Note that, since $V_{2}=0$ when Dirac brackets are used,
equations (2.12) and (2.14) can be written as
$$
\dot{\varphi}_{i} \approx r^{-4} p_{\varphi_{i}}
\eqno (2.15)
$$
$$
\dot{p}_{\varphi_{i}} \approx r^4
{\delta V \over \delta \varphi_{i}}~~~.
\eqno (2.16)
$$
For the purpose of numerical integration, the most convenient form
of these equations is
$$
{d \over d \tau} \left( r^4 \dot{\varphi}_{i} \right) \approx
r^4 {\delta V \over \delta \varphi_{i}}
\eqno (2.17)
$$
$$
\dot{r} \approx - {r^{3}\left(8 \sum_{i=1}^{5} \dot{\varphi}_{i}
{\delta V \over \delta \varphi_{i}}
+ r \sum_{i=1}^{5} \dot{\varphi}_{i}
{\delta^{2} V \over \delta \varphi_{i} \delta r} \right) \over
{\left(-10 r^2 \sum_{i=1}^{5} \dot{\varphi}_{i}^{2} -2 b + 12 r^2 V
+ r^4 {\delta^2  V \over \delta r^2}
+ 8 r^3 {\delta V \over \delta r}
\right)}}~~~.
\eqno (2.18)
$$
Such a system is here solved choosing the following initial conditions:
$$
r(0) \equiv r_{0}= \sqrt{ 3/ 8 \pi G V_0 }
\eqno (2.19)
$$
$$
\varphi_{i}(0)= \varphi_{i}^{0}
\eqno (2.20)
$$
$$
p_{r}(0)=0
\eqno (2.21)
$$
$$
p_{\varphi_{i}}(0)=0
\eqno (2.22)
$$
where in equation (2.19) the value chosen for $r_{0}$
leads to a suitable cosmological
constant for the inflationary era, and in equation (2.22)
we have neglected for simplicity initial-kinetic-energy effects
(cf end of section 5).
Of course, the $\varphi_{i}^{0}$ values should obey the constraint
$V_{2}(\tau=0)=0$, i.e.
$$
0= -2br_{0} + 4 (r_{0})^{3} V(r_{0},{\hat {\bf \Phi}}_{0}) + (r_{0})^{4}
\left. {\delta V \over \delta r}
\right|_{r_{0},{\hat {\bf \Phi}}_{0}}
\; \; \; \; .
\eqno (2.23)
$$
{}From now on it is useful to use dimensionless units. For this purpose,
we define $\sigma \equiv \sqrt{2 G/ 3 \pi }=
\sqrt{2/(3 \pi M_{P}^2)}$ and make the rescalings
$r \rightarrow \sigma~ r$, $\tau \rightarrow \sigma~ \tau$ ,
$\varphi_{i} \rightarrow \phi_{i} /(\pi \sigma \sqrt{2})$,
and we also define
$2 \pi^{2} \sigma^{4} V(\sigma r,\phi_{i}/(\pi \sigma \sqrt{2})) \equiv
\widetilde{V}(r, \phi_{i})$. The dimensionless field equations
corresponding to equations (2.17)-(2.18) are then found to be
$$
{d \over d \tau} \left( r^4 \dot{\phi}_{i} \right) \approx
r^4 {\delta \widetilde{V} \over \delta \phi_{i}}
\eqno (2.24)
$$
$$
\dot{r} \approx - {r^{3}\left(8 \sum_{i=1}^{5} \dot{\phi}_{i}
{\delta \widetilde{V} \over \delta \phi_{i}} +
r \sum_{i=1}^{5} \dot{\phi}_{i}
{\delta^{2} {\widetilde V} \over
\delta \phi_{i} \delta r} \right) \over
\left(-10 r^2 \sum_{i=1}^{5} \dot{\phi}_{i}^{2} -2+ 12 r^2
\widetilde{V} + r^4 {\delta^2  \widetilde{V} \over \delta r^2}
+ 8 r^3 {\delta \widetilde{V} \over \delta r} \right)}~~~.
\eqno (2.25)
$$
Moreover, the initial conditions here chosen take the form
$$
r(0) \equiv  r_{0}= \sqrt{{1 \over 2{\widetilde V}_{0}}}
\cong {3 M_{P}^{2} \over 4 M_{X}^{2}}
\eqno (2.26)
$$
$$
\phi_{i}(0)= \phi_{i}^{0}
\eqno (2.27)
$$
$$
p_{r}(0)=0
\eqno (2.28)
$$
$$
p_{\varphi_{i}}(0)=0
\eqno (2.29)
$$
where the $\phi_{i}^{0}$ values obey, for a given $r_{0}$ value,
the constraint
$$
0= -2 r_{0} + 4 (r_{0})^{3} \widetilde{V}(r_{0},\phi^{0}_{i}) + (r_{0})^{4}
\left. {\delta \widetilde{V} \over \delta r} \right|_{r_{0},\phi^{0}_{i}}
\; \; \; \; .
\eqno (2.30)
$$
Note that the choice $V_{0}=M_{X}^{4}$ fixes reasonably the critical
temperature for the phase transitions to be of order $T_{C}
\cong M_{X}$ ($M_{X}$, approximately equal to $10^{15}$ Gev, is
the typical order of magnitude of the unification mass in the
minimal $SU(5)$ model).
\vskip 10cm
\leftline {\bf 3. $1$-loop effective potential}
\vskip 1cm
\noindent
We here study ${\bf \hat{\phi}}$
in the forms invariant under the subgroups
$SU(3) \times SU(2) \times U(1)$ and $SU(4) \times U(1)$
respectively, i.e.
$$
{\bf \hat{\phi}_{321}} = {\| {\bf \hat{\phi}_{321} }\|^{1/2}\over
\sqrt{30}} {\rm diag}(2,2,2,-3,-3)
\eqno (3.1)
$$
$$
{\bf \hat{\phi}_{41}} = {\| {\bf \hat{\phi}_{41} }\|^{1/2} \over
\sqrt{20}} {\rm diag}(1,1,1,1,-4)
\eqno (3.2)
$$
since it has been shown in Allen 1985 and Buccella {\it et al} 1992
that these are the only subgroups relevant to the $SU(5)$
symmetry-breaking pattern. From now on, we denote by $\gamma(\tau)$
the norm of ${\bf \hat{\phi}_{321}}$ or of
${\bf \hat{\phi}_{41}}$, which is the only variable characterizing
these broken-symmetry phases.

The form of the dimensionless effective potential in the $321$
and $41$ directions (i.e. when ${\bf {\hat \phi}}=
{\bf \hat{\phi}_{321}}$ or ${\bf {\hat \phi}}=
{\bf \hat{\phi}_{41}}$) is obtained inserting equations (3.1)-(3.2)
into equation (1.1), which yields
$$ \eqalignno{
\widetilde{V}(r,{\bf
\hat{\phi}_{321}})|_{\| {\bf \hat{\phi}}\|^{1/2} = \gamma}
& \equiv \widetilde{V}_{321}(r,\gamma)\cr
&={45 \alpha \over 4 \pi r^{2}}
\left \{ Q+{1\over 3} \left[ 1-\log \biggr({2 M_{X}^{2}\over
3 \pi M_{P}^{2}} r^{2} \biggr) \right] \right \}\gamma^{2}\cr
&+ {25 \alpha^{2} \over 32 \pi^{2}}
\left[ 1-\log \biggr({2 M_{X}^{2} \over
3 \pi M_{P}^{2}} r^{2} \biggr) \right] \gamma^{4}\cr
&-{9\over 2 r^{4}}
{\cal A}\left[{5 \alpha  \over 6 \pi} r^{2}
{\gamma}^{2} \right] + \widetilde{V}_{0}
&(3.3)\cr}
$$
$$ \eqalignno{
\widetilde{V}(r,{\bf \hat{\phi}_{41}})|_{\| {\bf \hat{\phi}}\|^{1/2} = \gamma}
& \equiv \widetilde{V}_{41}(r,\gamma)\cr
&={45 \alpha \over 4 \pi r^{2}}
\left \{ Q+{1\over 3} \left[ 1-\log \biggr({2 M_{X}^{2} \over
3 \pi M_{P}^{2}} r^{2}\biggr) \right] \right\} \gamma^{2}\cr
&+{75 \alpha^{2} \over 16 \pi^{2}} \left\{ {1\over 4}
\left[ 1-\log \biggr({2 M_{X}^{2} \over
3 \pi M_{P}^{2}} r^{2} \biggr) \right] +
{\widetilde{\Lambda} \over 5} \right\} \gamma^{4}\cr
&-{3\over r^{4}}
{\cal A}\left[{5 \alpha  \over 4 \pi} r^{2}
{\gamma}^{2} \right] + \widetilde{V}_{0}
&(3.4)\cr}
$$
where $\widetilde{V}_{0} \equiv (8 M_{X}^{4})/(9 M_{P}^{4})$ and $\alpha
\equiv g^{2}/4 \pi$. It is also useful to derive the
$r \rightarrow \infty$ limit of these potentials (i.e. their
flat-space limit) as
$$
\widetilde{V}({\bf \hat{\phi}_{321}})|_{\| {\bf \hat{\phi}}\|^{1/2} = \gamma}
(r \rightarrow \infty) = \widetilde{V_{0} }
+ {25 \alpha^{2} \over 32 \pi^{2}} \gamma^{4}
\left[ \log{\left({5 \alpha M_{P}^{2} \over 4 M_{X}^{2}}
\gamma^2 \right)} -{1 \over 2}
\right]
\eqno (3.5)
$$
$$
\widetilde{V}({\bf \hat{\phi}_{41}})|_{\| {\bf \hat{\phi}}\|^{1/2} = \gamma}
(r \rightarrow \infty) = \widetilde{V_{0}} +
{15 \alpha^{2} \over 16 \pi^2} \widetilde{\Lambda} \gamma^{4}
+ {75 \alpha^{2} \over 64 \pi^{2}} \gamma^{4}
\left[ \log{\left({15 \alpha M_{P}^{2} \over 8 M_{X}^{2}}
\gamma^2\right)} -{1 \over 2}  \right] \; \; \; \; .
\eqno (3.6)
$$
The potentials (3.5) and (3.6) evaluated at their minima, denoted
for simplicity by $\gamma_{m}$, turn out to be
$$
\widetilde{V}({\bf \hat{\phi}_{321}})|_{\| {\bf \hat{\phi}}\|^{1/2} =
\gamma_{m}}
(r \rightarrow \infty) = \widetilde{V}_{0} -
{1 \over 4 \pi^{2}}
{M_{X}^{4} \over M_{P}^{4}}
\eqno (3.7)
$$
$$
\widetilde{V}({\bf \hat{\phi}_{41}})|_{\| {\bf \hat{\phi}}\|^{1/2} =
\gamma_{m}}
(r \rightarrow \infty) = \widetilde{V}_{0} - {1 \over 6 \pi^{2}}
{M_{X}^{4} \over M_{P}^{4}}
\exp\left[ - {8 \over 5} \widetilde{\Lambda}\right]
\; \; \; \; .
\eqno (3.8)
$$
The experimental evidence for the $SU(3) \times SU(2) \times U(1)$
gauge symmetry at energy $E$ greater than or of order $100$ GeV
requires for the scalar potential
that the absolute minimum of $\widetilde{V}({\bf \hat{\phi}_{321}})
(r \rightarrow \infty)$ should remain below the absolute minimum of
$\widetilde{V}({\bf \hat{\phi}_{41}})(r \rightarrow \infty)$ and below
$\widetilde{V}_{0}$. One thus finds the condition on the bare parameters
$\Lambda_{2}$ and $\Lambda_{4}$
(Allen 1985) of the scalar potential
$$
{3 \over 5} \Lambda_{4} - \Lambda_{2} >
- {75 \alpha^{2} \over 32}
\log{\left({3 \over 2} \right)}
\; \; \; \; .
\eqno (3.9)
$$
{}From now on, we restrict the choice of $\Lambda_{2}$ and $\Lambda_{4}$
to the region defined by the inequality (3.9). Under the assumptions
described so far, the initial conditions (2.27) and (2.29) take the form
$$
\gamma(0)= \gamma_{0}
\eqno (3.10)
$$
$$
\dot{\gamma}(0) = 0~~~.
\eqno (3.11)
$$
The $\gamma_{0}$ value and the
corresponding residual symmetry are obtained
by solving separately the constraint equations (cf equation (2.23))
$$
0= -2 r_{0} + 4 (r_{0})^{3}
\widetilde{V}_{321}(r_{0},\gamma_{0}) + (r_{0})^{4}
\left. {\delta \widetilde{V}_{321}
\over \delta r} \right|_{r_{0},\gamma_{0}}
\eqno (3.12)
$$
$$
0= -2 r_{0} + 4 (r_{0})^{3}
\widetilde{V}_{41}(r_{0},\gamma_{0}) + (r_{0})^{4}
\left. {\delta \widetilde{V}_{41}
\over \delta r} \right|_{r_{0},\gamma_{0}}
{}~~~.
\eqno (3.13)
$$
After doing this, one compares the $\widetilde{V}_{321}(r_{0},\gamma_{0})$
and $\widetilde{V}_{41}(r_{0},\gamma_{0})$ values, requiring that the
correct initial condition should lead to the minimum value of the
effective potential. Once the correct initial condition has been picked
out in this way, the system (2.24)-(2.25) expressed in terms of
$\gamma$ becomes
$$
{d \over d \tau} \left( r^4 \dot{\gamma} \right) \approx
r^4 {\delta \widetilde{V} \over \delta \gamma}
\eqno (3.14)
$$
$$
\dot{r} \approx - {r^{3} {\dot \gamma}\left(8
{\delta \widetilde{V} \over \delta \gamma} +
r {\delta^{2} \widetilde{V} \over \delta \gamma \delta r}
\right) \over
\left(-10 r^2 \dot{\gamma}^{2} -2+ 12 r^2
\widetilde{V} + r^4 {\delta^2  \widetilde{V} \over \delta r^2}
+ 8 r^3 {\delta \widetilde{V} \over \delta r} \right)}
\eqno (3.15)
$$
jointly with the constraint (cf equation (2.6))
$$
V_{2}(\tau) \equiv 2 r^{3} {\dot{\gamma}}^2
-2 r + 4 r^{3} \widetilde{V}(r,\gamma) + r^{4}
{\delta \widetilde{V} \over \delta r}
\eqno (3.16)
$$
which should vanish $\forall \tau$.
\vskip 1cm
\leftline {\bf 4. Absolute minimum}
\vskip 1cm
\noindent
By virtue of equations (A.7)-(A.19) of the appendix, the asymptotic form
of the constraints (3.12)-(3.13) is
$$
{25 \alpha^2 \over 32 \pi^2 } \gamma^{4}_0 \biggr[
\log\biggr({5 \alpha M_{P}^{2} \over 4M_{X}^{2}} \gamma^{2}_0 \biggr)
-{1 \over 2} \biggr]  + {\rm O}\Bigr(r_{0}^{-2}\Bigr)= 0
\eqno (4.1)
$$
$$
{25 \alpha^2 \over 32 \pi^2 } \gamma^{4}_{0} \biggr[{3 \over 2}
\log\biggr({15 \alpha M_{P}^{2} \over 8M_{X}^{2}} \gamma^{2}_{0} \biggr)
-{3 \over 4} + {6 \over 5}{\widetilde \Lambda} \biggr]
+ {\rm O}\Bigr(r_{0}^{-2}\Bigr)
= 0
\; \; \; \; .
\eqno (4.2)
$$
The numerical solution of equation (4.1) yields
$\gamma_{0}\cong 1.42 \cdot 10^{-3}$, where the NAG-library routine C05ADF
has been used in double-precision. Such a value of $\gamma_{0}$ is
compatible with the asymptotic formulae appearing in the appendix.
Interestingly, if the inequality (3.9) holds, one finds
$\forall \tau$
$$
\widetilde{V}_{41}(\tau) -\widetilde{V}_{321}(\tau)
\sim {25 \alpha^2 \over 32 \pi^2}
\gamma^{4} \biggr[ {3 \over 2} \log\Bigr({3 \over 2}\Bigr) - { 1 \over 4}
+{6 \over 5} {\widetilde \Lambda} \biggr] >
{25 \alpha^{2} \over 32 \pi^{2}} \gamma^{4}
\biggr[{3 \over 4} \log\Bigr({3 \over 2}\Bigr) -
{1 \over 4}\biggr] > 0
\; \; \; \; .
\eqno (4.3)
$$
This result ensures that at $\tau=0$, if $r_{0}$ is taken to be of
order $10^{8}$, so that the asymptotic formulae of the appendix can
be applied, the only possible residual symmetry is
$SU(3) \times SU(2) \times U(1)$. Remarkably, because equation (4.1)
is, at least up to first order in the $r_{0}$-expansion,
independent of the bare parameters (${\widetilde \Lambda},Q$)
of the scalar potential and only dependent on
$M_{X}$, the value found for $\gamma_{0}$ is
basically model-independent. If the alternative symmetry
$SU(4) \times U(1)$ were chosen, the inequality (4.3) would lead to a
tunneling of the Higgs field towards the energetically more favourable
phase. Moreover, our result (4.3) ensures that, once the initial
$SU(3) \times SU(2) \times U(1)$ symmetry is chosen, the Higgs field
remains in this broken-symmetry phase during the whole de Sitter phase
of the early universe.
\vskip 1cm
\leftline {\bf 5. Numerical analysis}
\vskip 1cm
\noindent
The numerical integration of the system (3.14)-(3.15) can be performed
after reduction to first-order form, and using the second-class
constraint (3.16) and equations (2.15)-(2.16). Thus, defining
$$
\dot{\gamma}(\tau) \equiv \eta(\tau)
\eqno (5.1)
$$
this leads to
$$
\dot{\eta}(\tau) \approx {\delta \widetilde{V} \over \delta \gamma} -
{4 \eta^{2}r^{2}
\left(8 {\delta \widetilde{V} \over \delta \gamma} +
r {\delta^{2} \widetilde{V} \over \delta \gamma \delta r}\right)
\over
\left(12-32r^{2} \widetilde{V}
-13 r^{3} {\delta \widetilde {V} \over \delta r}
-r^{4} {\delta^{2} \widetilde {V} \over \delta r^{2}}\right)}
\eqno (5.2)
$$
$$
\dot{r}(\tau) \approx { r^{3}\eta \left(8
{\delta \widetilde{V} \over \delta \gamma} + r
{\delta^{2} \widetilde{V} \over \delta \gamma \delta r}\right) \over
\left(12-32r^{2} \widetilde{V}
-13 r^{3} {\delta \widetilde {V} \over \delta r}
-r^{4} {\delta^{2} \widetilde {V} \over \delta r^{2}}\right)}
\; \; \; \; .
\eqno (5.3)
$$
By taking $r_{0} \cong 10^{8}$, which is a typical value for GUT models
in a de Sitter universe, the system (5.1)-(5.3) has been solved
within the framework of the Runge-Kutta-Merson method
(NAG-library routine D02BAF).
The results of our numerical
analysis are shown in figures $1-3$, which present the
Euclidean-time evolution of $\gamma$, $\dot \gamma$ and $r$ respectively.
These plots clearly show the existence of
an {\it almost exact} de Sitter phase
(i.e. when the four-sphere radius remains approximately constant)
with typical time $\tau$, in dimensionless units, of order
$3 \cdot 10^{4}$,
whereas the duration of the exponentially-expanding phase
may be taken to be of order $7 \cdot 10^{4}$. Interestingly, this result
is practically independent of the bare parameters in the $SU(5)$ potential
we have chosen if the inequality (3.9) is satisfied. Further
restrictions are then given by particle physics, i.e. proton-lifetime
experiments and renormalization-group
equations for the coupling constants and $(\sin \theta_{W})^{2}$
(Buccella {\it et al} 1989,
Becker-Szendy {\it et al} 1990, Amaldi {\it et al} 1992).

If $\tau \geq 7 \cdot 10^{4}$, figure $3$ shows a rapid variation of
the four-sphere radius, so that the early universe is no longer
well described by a de Sitter or exponentially-expanding model.
Note also that the numerical analysis here presented rules out the
occurrence of tunneling effects between broken-symmetry phases.
Moreover, it should be emphasized that, setting to zero the initial
kinetic energy (cf equation (3.11)) one obtains the most favourable
initial conditions for a long de Sitter phase of the early universe,
whereas values of ${\dot \gamma}(0) \not = 0$ may be shown to lead
to a much more rapid variation of the four-sphere radius $r$.
\vskip 1cm
\leftline {\bf 6. Concluding remarks}
\vskip 1cm
\noindent
This paper has studied the $SU(5)$ symmetry-breaking pattern in
a de Sitter universe from a Hamiltonian point of view. As a result
of this analysis, one finds that the model is characterized by two
second-class constraints. Interestingly, the secondary
second-class constraint has been used to prove that the early
universe can only reach the $SU(3) \times SU(2) \times U(1)$
broken-symmetry phase if we require
the correct low-energy limit for the GUT
theory. Furthermore, under this obvious requirement,
the $SU(3) \times SU(2) \times U(1)$ invariant direction
turns out to be energetically more favourable than $SU(4) \times U(1)$.
Thus, during the whole inflationary era, possible tunneling effects
between the two broken-symmetry phases are forbidden.

This conclusion supersedes earlier work on the $SU(5)$
symmetry-breaking pattern in de Sitter cosmologies appearing in
Allen 1985 and Buccella {\it et al} 1992. It also provides a relevant
example of cosmological restrictions on GUT models (cf Collins and
Langbein 1992).

Moreover, the resulting field equations have been solved
numerically. Standard methods for the numerical integration
of first-order systems of ordinary differential equations show
that the early universe starts out in a de Sitter state. The exact
de Sitter state is then replaced by a more general
exponentially-expanding universe corresponding to the
slow-rolling-over phase, as one would expect. In dimensionless
units, the total duration of these two phases has been found to
be of order $7 \cdot 10^{4}$.

{}From a field-theoretical point of view, the $1$-loop effective
potential of equation (1.1), originally derived in Allen 1985,
and also used in Buccella {\it et al} 1992 and in this paper,
might be subject to criticisms, since topological methods have
been used to prove the general non-existence of a gauge choice
for a theory with gauge group $SU(N)$, $N \geq 3$, over the
manifolds $S^{3}$ and $S^{4}$ (Jungman 1992 and references
therein). However, this difficulty holds for a classical
non-Abelian gauge-field theory. By contrast, we have studied
the quantization of a Yang-Mills-Higgs theory where
gauge-averaging terms are included in the Wick-rotated path
integral using the Faddeev-Popov technique and a specific choice
first proposed by 't Hooft (Allen 1985, Buccella {\it et al} 1992).
Although the choice of a gauge-averaging term is clearly suggested
by what would be set to zero at the classical level, there is by
now some evidence that the resulting quantum theory is not equivalent
to the model where only physical degrees of freedom are quantized
after fixing the gauge (Esposito 1992 and references therein). It
therefore appears that the approach used in Allen 1985 to evaluate
the $1$-loop effective potential for $SU(5)$ theory in de Sitter
cosmologies remains a useful tool to improve our understanding of
the symmetry-breaking pattern of non-Abelian gauge theories in
curved backgrounds.
\vskip 1cm
\leftline {\bf Appendix}
\vskip 1cm
\noindent
The function ${\cal A}(z)$ appearing in equation (1.1), and its
derivatives, are given by (Allen 1985)
$$ \eqalignno{
{\cal A}(z) & \equiv  {z^{2}\over 4} + {z \over 3} -
\int_{2}^{{3 \over 2}
+ \sqrt{ {1 \over 4} - z }} y \Bigr( y -{3 \over 2}
\Bigr) \Bigr( y - 3  \Bigr) \psi(y)~ dy \cr
&-\int_{1}^{{3 \over 2} - \sqrt
{ {1 \over 4} - z }} y \Bigr( y -{3 \over 2}
\Bigr) \Bigr( y - 3  \Bigr) \psi(y)~ dy
&(A.1)\cr}
$$
$$
{\cal A}'(z) = - {1 \over 2} (z+2) \left[ \psi\biggr(
{3 \over 2} + \sqrt{ {1 \over 4} - z } \biggr) +
\psi\biggr(
{3 \over 2} - \sqrt{ {1 \over 4} - z } \biggr) \right]
+{z \over 2} + {1 \over 3}
\eqno (A.2)
$$
$$ \eqalignno{
{\cal A}''(z) & =  {1 \over 2} - {1 \over 2} \left[ \psi\biggr(
{3 \over 2} + \sqrt{ {1 \over 4} - z } \biggr) +
\psi\biggr({3 \over 2} - \sqrt{ {1 \over 4} - z } \biggr) \right]\cr
&+ { 1 \over 4} {(z+2) \over \sqrt{\left( {1 \over 4}- z \right)}}
\left[ \psi'\biggr(
{3 \over 2} + \sqrt{ {1 \over 4} - z } \biggr) -
\psi'\biggr({3 \over 2} - \sqrt{ {1 \over 4} - z} \biggr) \right]~~~.
&(A.3)\cr}
$$
At large $z$, the following asymptotic expansions hold (Allen 1985):
$$
{\cal A}(z) \sim - \left( {z^2 \over 4} + z + {19 \over 30} \right)
\log{(z)} + {3 \over 8}z^2 + z
+ {\rm const.} + {\rm O}\left(z^{-1}\right)
\eqno (A.4)
$$
$$
{\cal A}'(z) \sim - {1 \over 2}(z+2) \log{(z)} + {z \over 2} +
{\rm O}\left(z^{-1}\right)
\eqno (A.5)
$$
$$
{\cal A}''(z) \sim - {1 \over 2}\log{(z)}+ {\rm O}\left(z^{-1}\right)
\; \; \; \; .
\eqno (A.6)
$$
In section 4, we rely on the following exact formulae for the
derivatives of the dimensionless effective potential in the
$SU(3) \times SU(2) \times U(1)$ and $SU(4) \times U(1)$
broken-symmetry phases:
$$ \eqalignno{
{\delta \widetilde{V}_{321} \over \delta r}(r,\gamma)
&=-{45 \alpha \over 2 \pi r^{3}}
\left \{ Q+{1\over 3} \biggr[2-\log \biggr({2 M_{X}^{2}\over
3 \pi M_{P}^{2}} r^{2} \biggr) \biggr] \right \}\gamma^{2}
- {25 \alpha^{2} \over 16 \pi^{2} r} \gamma^{4}\cr
&+{18\over r^{5} }
{\cal A}\left[{5 \alpha  \over 6 \pi} r^{2}
{\gamma}^{2} \right] - {15 \alpha \over 2  \pi r^{3} } \gamma^{2}
{\cal A}'\left[{5 \alpha  \over 6 \pi} r^{2} {\gamma}^{2} \right]
&(A.7)\cr}
$$
$$ \eqalignno{
{\delta^{2} \widetilde{V}_{321} \over \delta r^{2}}(r,\gamma)
&={135 \alpha \over 2 \pi r^{4}}
\left \{ Q+{1\over 3} \biggr[{ 8 \over 3}-\log \biggr({2M_{X}^{2}\over
3 \pi M_{P}^{2}} r^{2} \biggr) \biggr] \right \}\gamma^{2}
+ {25 \alpha^{2} \over 16 \pi^{2} r^{2}} \gamma^{4}\cr
&-{90\over r^{6} }
{\cal A}\left[{5 \alpha  \over 6 \pi} r^{2}
{\gamma}^{2} \right] + {105 \alpha \over 2  \pi r^{4} } \gamma^{2}
{\cal A}'\left[{5 \alpha  \over 6 \pi} r^{2} {\gamma}^{2} \right]\cr
&-{25 \alpha^{2} \over 2  \pi^{2} r^{2} } \gamma^{4}
{\cal A}''\left[{5 \alpha  \over 6 \pi} r^{2} {\gamma}^{2} \right]
&(A.8)\cr}
$$
$$ \eqalignno{
{\delta^{2} \widetilde{V}_{321} \over \delta \gamma \delta r}(r,\gamma)
&=-{15 \alpha \over  \pi r^{3}}
\left \{ 3 Q+2 - \log \biggr({2 M_{X}^{2} \over
3 \pi M_{P}^{2}} r^{2} \biggr) \right \}\gamma
-{25 \alpha^{2} \over 4 \pi^{2} r } \gamma^{3}\cr
&+{15 \alpha \over \pi r^{3} } \gamma
{\cal A}'\left[{5 \alpha  \over 6 \pi} r^{2} {\gamma}^{2} \right]
-{25 \alpha^{2} \over 2  \pi^{2} r } \gamma^{3}
{\cal A}''\left[{5 \alpha  \over 6 \pi} r^{2} {\gamma}^{2} \right]
&(A.9)\cr}
$$
$$ \eqalignno{
{\delta \widetilde{V}_{321} \over \delta \gamma}(r,\gamma)
&={45 \alpha \over 2 \pi r^{2}}
\left \{ Q+{1\over 3} \biggr[1-\log \biggr({2M_{X}^{2} \over
3 \pi M_{P}^{2}} r^{2} \biggr) \biggr] \right \}\gamma \cr
&+{25 \alpha^2 \over 8 \pi^2 }
\biggr[1-\log \biggr({2M_{X}^{2} \over
3 \pi M_{P}^{2}} r^{2} \biggr) \biggr] \gamma^{3}
-{15 \alpha \over 2  \pi r^{2} } \gamma
{\cal A}'\left[{5 \alpha  \over 6 \pi} r^{2} {\gamma}^{2} \right]
&(A.10)\cr}
$$
$$ \eqalignno{
{\delta \widetilde{V}_{41} \over \delta r}(r,\gamma)
&=-{45 \alpha \over 2 \pi r^{3}}
\left \{ Q+{1\over 3} \biggr[ 2-\log \biggr({2M_{X}^{2} \over
3 \pi M_{P}^{2}} r^{2} \biggr) \biggr] \right \}\gamma^{2}
-{75 \alpha^{2} \over 32 \pi^{2} r} \gamma^{4} \cr
&+{12\over r^{5} }
{\cal A}\left[{5 \alpha \over 4 \pi} r^{2}
{\gamma}^{2} \right] - {15 \alpha \over 2  \pi r^{3} } \gamma^{2}
{\cal A}'\left[{5 \alpha  \over 4 \pi} r^{2} {\gamma}^{2} \right]
&(A.11)\cr}
$$
$$ \eqalignno{
{\delta \widetilde{V}_{41} \over \delta \gamma}(r,\gamma)
&={45 \alpha \over 2 \pi r^{2}}
\left \{ Q+{1\over 3} \biggr[ 1-\log \biggr({2M_{X}^{2} \over
3 \pi M_{P}^{2}} r^{2} \biggr) \biggr] \right \}\gamma \cr
&+{75 \alpha^{2} \over 4 \pi^{2}} \left\{ {1 \over 4}
\biggr[ 1-\log \biggr({2M_{X}^{2} \over
3 \pi M_{P}^{2}} r^{2} \biggr) \biggr] +
{\widetilde{\Lambda} \over 5} \right\} \gamma^{3} \cr
&-{15 \alpha \over 2 \pi r^{2} } \gamma
{\cal A}'\left[{5 \alpha \over 4 \pi} r^{2} {\gamma}^{2} \right]
\; \; \; \; .
&(A.12)\cr}
$$
In light of equations (A.4)-(A.6), the asymptotic expansions of the
dimensionless effective potential
and of equations (A.7)-(A.12) are given by
$$ \eqalignno{
\widetilde{V}_{321}(r,\gamma) & \sim \widetilde{V}_{0} +
{25 \alpha^2 \over 32 \pi^2 } \gamma^{4}\left[
\log\biggr({5 \alpha M_{P}^{2} \over 4 M_{X}^{2}} \gamma^{2}\biggr) -
{1 \over 2} \right] \cr
&+{15 \alpha \over 4 \pi r^{2}}
\biggr[3Q+ \log \biggr({5 \alpha M_{P}^{2} \over
4M_{X}^{2}} \gamma^{2}\biggr) \biggr] \gamma^{2}
+{\rm O}(r^{-4})
&(A.13)\cr}
$$
$$
{\delta \widetilde{V}_{321} \over \delta r}(r,\gamma)
\sim - {15 \alpha \over 2 \pi r^{3}}
\biggr[3Q+ \log \biggr({5 \alpha M_{P}^{2} \over
4M_{X}^{2}} \gamma^{2}\biggr) \biggr] \gamma^{2}
+{\rm O}(r^{-5})
\eqno (A.14)
$$
$$
{\delta^{2} \widetilde{V}_{321} \over \delta r^{2}}(r,\gamma)
\sim {135 \alpha \over 2 \pi r^{4}}
\left\{Q+ {1 \over 3} \biggr[- {2 \over 3} +
\log \biggr({5 \alpha M_{P}^{2} \over
4M_{X}^{2}} \gamma^{2}\biggr) \biggr] \right\}\gamma^2 +{\rm O}(r^{-6})
\eqno (A.15)
$$
$$
{\delta^{2} \widetilde{V}_{321} \over \delta \gamma \delta r}(r,\gamma)
\sim - {15 \alpha \over  \pi r^{3}}
\biggr[3Q+2 + \log \biggr({5 \alpha M_{P}^{2} \over
4M_{X}^{2}} \gamma^{2}\biggr) \biggr] \gamma +{\rm O}(r^{-5})
\eqno (A.16)
$$
$$ \eqalignno{
{\delta \widetilde{V}_{321} \over \delta \gamma}(r,\gamma)
& \sim {25 \alpha^2 \over 8 \pi^2 } \gamma^{3}
\log\biggr({5 \alpha M_{P}^{2} \over 4M_{X}^{2}} \gamma^{2}\biggr)\cr
&+{15 \alpha \over 2 \pi r^{2}}
\biggr[3Q+ 1+\log \biggr({5 \alpha M_{P}^{2} \over
4M_{X}^{2}} \gamma^{2}\biggr) \biggr] \gamma
+{\rm O}(r^{-4})
&(A.17)\cr}
$$
$$
{\delta \widetilde{V}_{41} \over \delta r}(r,\gamma)
\sim - {15 \alpha \over 2 \pi r^{3}}
\biggr[3 Q+ \log \biggr({15 \alpha M_{P}^{2} \over
8M_{X}^{2}} \gamma^{2}\biggr) \biggr] \gamma^{2} +{\rm O}(r^{-5})
\eqno (A.18)
$$
$$
{\delta \widetilde{V}_{41} \over \delta \gamma}(r,\gamma)
\sim {15 \alpha^2 \over 4 \pi^2 } \gamma^{3}\left\{ {5 \over 4}
\log\biggr({15 \alpha M_{P}^{2} \over 8M_{X}^{2}} \gamma^{2}\biggr)
+\widetilde{\Lambda} \right\} +{\rm O}(r^{-2})
\; \; \; \; .
\eqno (A.19)
$$
\vskip 10cm
\leftline {\bf Acknowledgments}
\vskip 1cm
\noindent
We are much indebted to Professor Franco Buccella for suggesting
the problem and reading the manuscript, and to
Professor Bruce Allen and Dr. Ya'qub
Anini for enlightening conversations.
The first author is grateful to Professor Abdus Salam,
the International Atomic Energy Agency and UNESCO for
hospitality at the International Centre for Theoretical
Physics, and to Professor Dennis Sciama for hospitality
at the SISSA of Trieste. Last but not least,
the second author has been financially supported
by the Istituto Nazionale di Fisica Nucleare
during his visit at the ICTP.
\vskip 1cm
\leftline {\bf References}
\vskip 1cm
\parindent=0pt
\everypar{\hangindent=20pt \hangafter=1}

Allen B 1983 {\it Nucl. Phys.} B {\bf 226} 228

Allen B 1985 {\it Ann. Phys.} {\bf 161} 152

Amaldi U, de Boer W, Frampton P H, F\"{u}rstenau H and Liu J T
1992 {\it Phys. Lett.} {\bf 281B} 374

Becker-Szendy R {\it et al} 1990 {\it Phys. Rev.} D {\bf 42} 2974

Buccella F, Ruegg H and Savoy C A 1980 {\it Nucl. Phys.}
B {\bf 169} 68

Buccella F, Miele G, Rosa L, Santorelli P and Tuzi T 1989
{\it Phys. Lett.} {\bf 233B} 178

Buccella F, Esposito G and Miele G 1992 {\it Class. Quantum Grav.}
{\bf 9} 1499

Collins P D B and Langbein R F 1992 {\it Phys. Rev.} D {\bf 45} 3429

Dirac P A M 1964 {\it Lectures on Quantum Mechanics}, Belfer Graduate
School of Science (New York: Yeshiva University)

Esposito G 1992 {\it Quantum Gravity, Quantum Cosmology and Lorentzian
Geometries}, Lecture Notes Series in Physics, New Series m: Monographs,
Volume m12 (Berlin: Springer-Verlag)

Jungman G 1992 {\it Mod. Phys. Lett.} A {\bf 7} 849
\vskip 100cm
\leftline {\bf Figure captions}
\vskip 1cm
\noindent
{\bf Figure 1.} The Euclidean-time evolution of the norm $\gamma$ of the
Higgs field is here plotted, after solving numerically the system
(5.1)-(5.3).
\vskip 1cm
\noindent
{\bf Figure 2.} The evolution of $\dot \gamma$ is here shown.
\vskip 1cm
\noindent
{\bf Figure 3.} The numerical solution for the Euclidean-time evolution
of the four-sphere radius $r$ is here presented. If
$\tau \in \Bigr[0,3 \cdot 10^{4}\Bigr]$, our solution of the system
(5.1)-(5.3) shows that $r$ is approximately constant, as one would
expect during the de Sitter phase of the early universe.
\bye